\documentclass[final,twocolumn]{IEEEtran}
\usepackage{graphicx,amssymb,amsmath,float,caption,tikz,booktabs,cite,paralist,epstopdf,setspace}
\usepackage[disable]{todonotes}

\long\def\comment#1{}

\newfont{\bbb}{msbm10 scaled 700}

\newfont{\bb}{msbm10 scaled 1100}

\newcommand{\EE}{\mbox{\bb E}}

\newcommand{\cv}{{\bf c}}

\newcommand{\hv}{{\bf h}}

\newcommand{\wv}{{\bf w}}

\newcommand{\xv}{{\bf x}}
\newcommand{\yv}{{\bf y}}
\newcommand{\zv}{{\bf z}}
\newcommand{\zerov}{{\bf 0}}

\newcommand{\Dm}{{\bf D}}

\newcommand{\Fm}{{\bf F}}

\newcommand{\Hm}{{\bf H}}
\newcommand{\Id}{{\bf I}}

\newcommand{\Km}{{\bf K}}

\newcommand{\Pm}{{\bf P}}

\newcommand{\Rm}{{\bf R}}
\newcommand{\Sm}{{\bf S}}

\newcommand{\Xm}{{\bf X}}

\newcommand{\Ac}{{\cal A}}

\newcommand{\Cc}{{\cal C}}
\newcommand{\Dc}{{\cal D}}

\newcommand{\Ic}{{\cal I}}

\newcommand{\Nc}{{\cal N}}
\newcommand{\Oc}{{\cal O}}

\newcommand{\Sc}{{\cal S}}

\newcommand{\Xc}{{\cal X}}
\newcommand{\Yc}{{\cal Y}}
\newcommand{\Zc}{{\cal Z}}

\newcommand{\Phim}{\hbox{\boldmath$\Phi$}}

\newcommand{\Psim}{\hbox{\boldmath$\Psi$}}

\newcommand{\Thetam}{\hbox{\boldmath$\Theta$}}

\newcommand{\diag}{{\hbox{diag}}}

\renewcommand{\arg}{{\hbox{arg}}}

\renewcommand{\Re}{{\rm Re}}
\renewcommand{\Im}{{\rm Im}}

\newcommand{\herm}{{\sf H}}

\newcommand{\transp}{{\sf T}}

\newcommand{\Afd}{\mbox{$\boldsymbol{\mathcal{A}}$}}

\newcommand{\Cfd}{\mbox{$\boldsymbol{\mathcal{C}}$}}
\newcommand{\Dfd}{\mbox{$\boldsymbol{\mathcal{D}}$}}

\newcommand{\Ufd}{\mbox{$\boldsymbol{\mathcal{U}}$}}

\newcommand{\Xfd}{\mbox{$\boldsymbol{\mathcal{X}}$}}
\newcommand{\Yfd}{\mbox{$\boldsymbol{\mathcal{Y}}$}}
\newcommand{\Zfd}{\mbox{$\boldsymbol{\mathcal{Z}}$}}

\begin{document}
\title{Receiver-based Recovery of Clipped OFDM Signals for PAPR Reduction: A Bayesian Approach}
\author{Anum Ali, {\it Student Member, IEEE}, Abdullatif Al-Rabah, Mudassir Masood, {\it Student Member, IEEE} \\and Tareq Y. Al-Naffouri$^{\ast}$, {\it Member, IEEE}
\thanks{The authors are with the department of Electrical Engineering, King Abdullah University of Science and Technology (KAUST), Thuwal, Saudi Arabia. T. Y. Al-Naffouri is also associated with the department of Electrical Engineering, King Fahd University of Petroleum and Minerals (KFUPM), Dhahran, Saudi Arabia.}%
\thanks{Corresponding Author: T. Y. Al-Naffouri (tareq.alnaffouri@kaust.edu.sa).}
\thanks{This work was supported in part by KAUST and in part by King Abdulaziz City for Science and Technology (KACST) through the Science \& Technology Unit at KFUPM, under Project No. $11$-ELE$1651$-$04$, as part of the National Science, Technology and Innovation Plan.}
\thanks{Part of this work was presented at European Signal Processing Conference (EUSIPCO), 2013 \cite{alrabah2013} and International Conference on Acoustics, Speech, and Signal Processing (ICASSP), 2014 \cite{ali2014receiver}.}
}
\maketitle
\begin{abstract}
Clipping is one of the simplest peak-to-average power ratio (PAPR) reduction schemes for orthogonal frequency division multiplexing (OFDM). Deliberately clipping the transmission signal degrades system performance, and clipping mitigation is required at the receiver for information restoration. In this work, we acknowledge the sparse nature of the clipping signal and propose a low-complexity Bayesian clipping estimation scheme. The proposed scheme utilizes \emph{a priori} information about the sparsity rate and noise variance for enhanced recovery. At the same time, the proposed scheme is robust against inaccurate estimates of the clipping signal statistics. The undistorted phase property of the clipped signal, as well as the clipping likelihood, is utilized for enhanced reconstruction. Further, motivated by the nature of modern OFDM-based communication systems, we extend our clipping reconstruction approach to multiple antenna receivers, and multi-user OFDM. We also address the problem of channel estimation from pilots contaminated by the clipping distortion. Numerical findings are presented, that depict favourable results for the proposed scheme compared to the established sparse reconstruction schemes.
\end{abstract}
\section{Introduction}\label{secIntro}
\IEEEPARstart{T}{he} problem of high peak-to-average power ratio (PAPR) in orthogonal frequency division multiplexing (OFDM) has received considerable research interest in the past. As the power amplifiers (PA) have a nonlinear response for higher input levels, inflated PAPR causes nonlinear distortion. Though power back-off in the operating point of the PA will reduce the nonlinear distortion, it is not desirable as it results in inefficient operation of the PA and reduced battery life of the mobile terminal. Hence PAPR reduction in OFDM signalling is a necessity for the linear and power efficient operation of the PA. Some of the transmitter-based PAPR reduction schemes include coding, partial transmit sequence (PTS), selected mapping (SLM), interleaving, tone reservation (TR), tone injection (TI) and active constellation extension (ACE)~\cite{han2005overview,wang2006overview,jiang2008overview,lim2009overview,tarokh2000computation}. The computational requirements of the aforementioned schemes make them unsuitable for applications where the transmitter complexity is a bottleneck, especially when the number of sub-carriers is large \cite{ochiai2002performance}.

Clipping is one simple and low-complexity PAPR reduction method. The clipping operation is performed such that the magnitude of the time-domain OFDM signal be limited to a pre-specified threshold. The clipping operation, however, is nonlinear and causes out-of-band radiation as well as in-band distortion. The out-of-band power spill interferes with adjacent channels and reduces power spectral efficiency. Though filtering can be used to significantly reduce the out-of-band radiation, it results in peak regrowth. A compromise between out-of-band spill and peak regrowth can be reached by iterative clipping and filtering operations \cite{armstrong2002peak,wang2005simplified,wang2011optimized,deng2007recursive}. Unlike out-of-band radiation, the in-band distortion can be taken care of at the receiver. However, if not, it results in significant performance loss evidenced e.g., by the high bit error rate (BER).


Recently the sparsity of the clipping signal has been exploited and compressed sensing (CS) schemes were used for clipping recovery at the receiver. The sparse nature of the clipping signal is evident as it originates when a high PAPR signal (with only a few high peaks) is subjected to a thresholding operation. Here it is noteworthy that the performance and applicability of any CS-based PAPR reduction scheme is mainly limited by two factors: the complexity of the sparse signal reconstruction scheme and the number of reserved/measurement tones. In \cite{al2012peak} Al-Safadi and Al-Naffouri utilized augmented CS for signal recovery in severe clipping scenarios. However, the drawback of \cite{al2012peak} is the severe hit taken on the data rate due to dedicated measurement tones. A CS-based approach using reliable carriers (RC) as measurement tones with no compromise on data rate is proposed in \cite{al2012pilotless}. However, this method is tailored for one-user, single-input-single-output systems and lacks the generality required by multiple-receiver antenna systems and multi-user communications.

In this work, we focus on deliberate clipping-based PAPR reduction. The time-domain OFDM signal is limited to a pre-specified threshold and the sparse clipping signal is reconstructed at the receiver using a low-complexity Bayesian recovery algorithm. The proposed reconstruction scheme is agnostic to the signal statistics and utilizes \emph{a priori} information about the additive noise, the sparsity rate of the signal and the clipping threshold. However, if accurate estimates of these quantities are not available, it can bootstrap itself and estimate them from the data. The proposed scheme also utilizes the \emph{a priori} information about the undistorted phase of the clipped signal for enhanced recovery. Further, the recovery algorithm focuses on the most probable clipping locations by obtaining the clipping likelihood from a comparison between the magnitude of the received data samples and the clipping threshold. At the receiver, some of the data sub-carriers are designated as RCs for sensing the clipping distortion (based on the criteria proposed in \cite{al2012pilotless}) and hence there is no data loss in using the proposed clipping reconstruction scheme. Considering that most modern communication standards use multiple antennas at the receiver, the proposed scheme is extended to the case of single-input multiple-output (SIMO) systems. It is also highlighted that the problem of clipping estimation in multi-user communication (i.e., orthogonal frequency division multiple access (OFDMA) systems) is not straightforward. The complications arise due to the fact that clipping distortions from different users overlap in frequency-domain and are indistinguishable from one another. In light of this, a clipping reconstruction scheme for OFDMA systems is also framed. The proposed multi-user clipping recovery scheme initially performs joint estimation of clipping distortion from all users. This is followed by the decoupling stage, in which subsystems belonging to each user are formed such that they are interference free from other users' distortion. Then the clipping is individually recovered on each decoupled subsystem. Lastly, we consider the channel estimation problem for clipped OFDM and present two data-aided channel estimation schemes. The main idea is to use RCs in addition to the pilot sub-carriers for enhanced minimum mean square error (MMSE) estimation.
\subsection{Notation}
Unless otherwise noted, scalars are represented by italic letters (e.g., $N$). Bold-face lower-case letters (e.g., $\xv$) are reserved to denote time-domain vectors, whereas frequency-domain vectors are represented using bold-face upper-case calligraphic letters (e.g., $\Xfd$). Bold-face upper-case letters are associated with matrices (e.g., $\Xm$). The symbols $\hat\xv$, $x(i)$, $\xv^\transp$ and $\xv^\herm$ respectively represent the estimate, $i{\rm th}$ entry, transpose and Hermitian (conjugate transpose) of the vector $\xv$. The operator $|\cdot|$ operating on a scalar (e.g., $|x(i)|$) will give the absolute value whereas operating on a set (e.g., $|\Sc|$) will give the number of elements in $\Sc$. Further, $\EE[\cdot]$, $\Id$ and $\zerov$ denote the expectation operator, identity matrix and the zero vector respectively. The operator $\diag(\Xm)$ forms a column vector $\xv$ from the diagonal of $\Xm$ and $\diag(\xv)$ constructs a diagonal matrix $\Xm$ with $\xv$ on its diagonal. Finally, $\underline{\Xfd}^u$ represents the $u{\rm th}$ portion of the vector $\Xfd$, where $\Xfd$ is partitioned in $U$ segments.
\subsection{Key Contributions}
The main contributions of this work can be summarized as follows
\begin{itemize}
\item A low-complexity Bayesian clipping recovery scheme is presented, that has the following features
    \begin{itemize}
    \item It is agnostic to the signal statistics.
    \item It uses \emph{a priori} information about the additive noise, the sparsity rate and the threshold. Further, it can bootstrap itself if accurate estimates of these parameters are not available.
    \item It utilizes the \emph{a priori} information of the undistorted phase and the clipping likelihood.
    \item It has a data-aided version that makes use of the RCs in place of reserved sub-carriers hence conserving the data rate.
    \end{itemize}
\item It is able to make use of the multiple receive antennas for enhanced clipping recovery.
\item It can be extended to the multi-user OFDMA systems.
\end{itemize}
In addition, this paper proposes effective channel estimation strategies that work in spite of pilot contamination from clipping distortion.
\subsection{Paper Organization}
The remainder of the paper is organized as follows. Section~\ref{secDatMod} introduces the data model for clipped OFDM signals and Section~\ref{secPCRS} formulates the proposed Bayesian clipping reconstruction scheme. The proposed scheme is then extended to SIMO systems in Section~\ref{secSIMOS}. A multi-user clipping recovery scheme is outlined in Section~\ref{secMUComm}. Section~\ref{secCEPC} presents the data-aided channel estimation strategies for clipped OFDM and Section~\ref{secConc} concludes the paper.
\section{Data Model for Clipped OFDM}\label{secDatMod}
In OFDM transmission, the incoming bitstream is first divided into $N$ parallel streams and is then modulated using an $M$-QAM constellation $\{\Ac_0,\Ac_1,\cdots,\Ac_{M-1}\}$. The modulated data $\Xfd=[\Xc(0),\Xc(1),\cdots,\Xc(N-1)]^\transp$, is converted to the time domain using the inverse discrete Fourier transform (IDFT) i.e., $\xv=\Fm^\herm\Xfd$. Here $\Fm$ is the DFT matrix whose ($n_1,n_2$) element is given by
\begin{equation}\nonumber
{\rm f}_{n_1,n_2}=N^{-1/2}e^{-\jmath2\pi n_1n_2/N},~~~n_1,n_2\in0,1,\cdots,N-1.
\end{equation}

The time-domain signal $\xv$ has a high PAPR and is subjected to an amplitude limiter for PAPR reduction. The resulting clipped signal $\xv_p$ is described by
\begin{align}\label{eqclip1}
 x_p(i) =
  \begin{cases}
   \gamma \exp(\jmath \angle{x(i)}) & \text{if}~~|x(i)| > \gamma \\
   x(i)       & \text{otherwise},
  \end{cases}
\end{align}
where $x_p(i)$ is the $i{\rm th}$ element of the signal after clipping, $\gamma$ is the limiting threshold and $\angle{x(i)}$ is the phase of $x(i)$. The clipping ratio ($\mathrm{CR}$) and threshold $\gamma$ are related by $\mathrm{CR}=\gamma/\sigma_x$, where $\sigma_x$ is the root mean squared power of the OFDM signal. The hard clipping in  (\ref{eqclip1}) is equivalent to the addition of a \emph{sparse} signal $\cv$ (with active elements only at the clipping locations) to the time-domain signal $\xv$. The clipped signal $\xv_p$ is then given as
\begin{align}\label{eqclip2}
\xv_p=\xv+\cv.
\end{align}
The equivalence of (\ref{eqclip1}) and (\ref{eqclip2}) dictates that the phase of $\cv$ must be the opposite of the phase of $\xv$ on the clipping locations and zero everywhere else. Thus, the addition of $\cv$ leaves the phase unaltered; i.e., $\angle{x_p(i)}=\angle{x(i)}=\angle{c(i)}+\pi~\forall~i$. This undistorted phase property is important and is exploited in the development of the proposed reconstruction scheme.

The clipped signal $\xv_p$ is transmitted through a channel of length $N_c$ with impulse response $\hv=[h(0),h(1),\cdots,h(N_c-1)]^\transp$, where the channel tap coefficients form a zero-mean complex Gaussian independent and identically distributed (i.i.d.) collection. The received time-domain signal can be written as
\begin{align}\label{eqrecsigtime}
\yv=\Hm\xv_p+\zv,
\end{align}
where $\Hm$ is the circulant channel matrix and $\zv$ is the additive white Gaussian noise (AWGN) with $\zv\sim\Cc\Nc(\zerov,\sigma_z^2\Id)$. The circulant nature of $\Hm$ allows us to diagonalize it using the DFT matrix $\Fm$ and write $\Hm=\Fm^\herm\Dm\Fm$, where $\Dm$ is a diagonal matrix with the channel frequency response on its diagonal. The data model and the proposed reconstruction scheme are developed assuming perfect channel knowledge at the receiver. The procedure for acquiring the channel impulse response (CIR) in clipped OFDM is outlined in Section~\ref{secCEPC}.

The frequency-domain received signal, obtained from (\ref{eqrecsigtime}) by the DFT operation, can be written as
\begin{align}\label{eqrecsigfreq}
\Yfd&=\Dm\Xfd_p+\Zfd=\Dm(\Xfd+\Cfd)+\Zfd,
\end{align}
where $\Yfd=\Fm\yv$ and $\Xfd_p,\Xfd,\Cfd,\Zfd$  are similarly defined. Equalizing the received data in (\ref{eqrecsigfreq}) results in
\begin{align}\label{eqrecsigfreqeq}
\widehat{\Xfd}=\Dm^{-1}\Yfd=\Xfd+\underbrace{\Cfd+\Dm^{-1}\Zfd}_{:=\Zfd^\dagger}=\Xfd+\Zfd^\dagger,
\end{align}
where $\Zfd^\dagger$ is the combined additive noise and clipping distortion. A na\"{i}ve receiver will disregard the presence of clipping noise in (\ref{eqrecsigfreqeq}) and will use maximum likelihood (ML) decoding on $\widehat{\Xfd}$ to obtain the estimated transmitted signal $\lfloor\widehat{\Xfd}\rfloor$ (the operator $\lfloor\cdot\rfloor$ is used to denote the ML decisions or equivalently rounding to the nearest QAM constellation point). However, a receiver employing CS reconstruction will exploit the sparse nature of $\cv$ for its estimation and hence removal.

As the clipping signal $\cv$ is sparse in the time-domain, its frequency-domain counterpart $\Cfd$ perturbs all sub-carriers alike as the time and frequency-domains are maximally incoherent. Utilizing this incoherence via CS, it is possible to reconstruct an $N$-dimensional time-domain sparse vector with only $P$ random projections on the frequency-domain, where $P<<N$. These projections can be made using randomly allocated pilot tones as in \cite{al2012peak}, but doing so reduces the data rate. In this work, we avoid this and use a data-aided approach to estimate $\cv$ as we describe below.

Given the equalized signal $\widehat{\Xfd}$ at the receiver, we expect the following: for some sub-carriers, the perturbation $\Zc^\dagger(i)$ is strong enough to take $\Xc(i)$ out of its correct decision region, i.e., $\lfloor\widehat{\Xc(i)}\rfloor\neq\Xc(i)$, while for others with a milder perturbation, we expect to have $\lfloor\widehat{\Xc(i)}\rfloor=\Xc(i)$. The subset of data sub-carries that satisfy $\lfloor\widehat{\Xc(i)}\rfloor=\Xc(i)$ are called RCs and fortunately constitute a major part of all sub-carriers. To select this subset, we note that the major source of perturbation is the clipping distortion, especially for high signal-to-noise ratio (SNR). Hence, from (\ref{eqrecsigfreqeq}), we can write the reliability function of the $i{\rm th}$ sub-carrier in terms of $\Zc^\dagger(i)$ as
\begin{align}\label{eqreliability}
\mathfrak{R}(i)=\dfrac{\mathrm{p}(\Zc^\dagger(i)=\widehat{\Xc(i)}-\lfloor\widehat{\Xc(i)}\rfloor)}{\sum_{k=0,\Ac(k)\neq\lfloor\widehat{\Xc(i)}\rfloor}^{M-1}\mathrm{p}(\Zc^\dagger(i)=\widehat{\Xc(i)}-\Ac(k))},
\end{align}
where $\mathrm{p}(\cdot)$ represents the pdf of $\Zfd^\dagger$, which is assumed to be zero mean Gaussian with variance $\sigma_z^2$ (see \cite{al2012pilotless} for details). In (\ref{eqreliability}), the numerator is the probability that $\Zc^\dagger(i)$ does not take $\Xc(i)$ beyond its correct decision region and the denominator sums the probabilities of all possible incorrect decisions that $\Zc^\dagger(i)$ can cause.

\begin{figure}[h!]\centering
\includegraphics[width=0.32\textwidth]{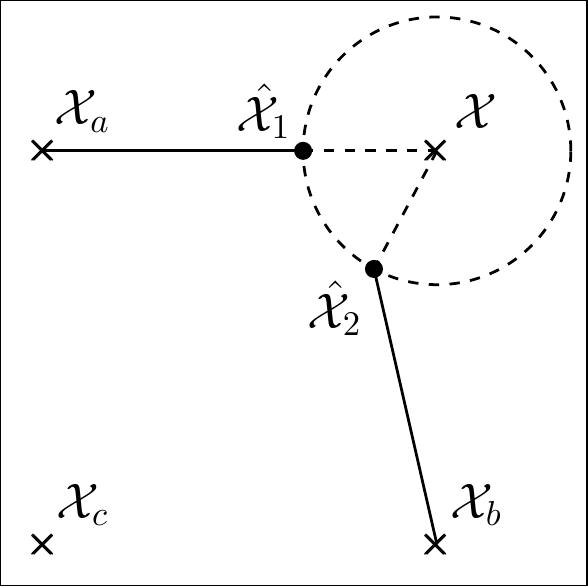}
\caption{Geometrical representation of the adopted reliability criteria.}
\label{da_fig}
\end{figure}
The utilized reliability criterion is unlike the Euclidean distance reliability criterion (employed in \cite{ali2013compressed}) that relies solely on the information of the distance between the received data point and the closest constellation point in $\Afd$. To this end, note that in Fig.~\ref{da_fig}, though $\hat\Xc_1$ and $\hat\Xc_2$ are equidistant from $\Xc$, they have different reliability values. This is owing to their distances with their next nearest neighbors, $\Xc_a$ and $\Xc_b$ respectively. The denominator in (\ref{eqreliability}) accounts for this fact, hence, yielding a more sophisticated reliability metric. The detailed investigation of this reliability criterion is reported in \cite{al2012pilotless}. After obtaining the reliability $\mathfrak{R}(i)$ for each carrier $i$, we pick the $P$ sub-carriers with the highest reliability values and use them as measurement tones to recover sparse clipping vector $\cv$. Consider an $N\times N$ binary selection matrix $\Sm$, with $P$ ones along its diagonal, corresponding to the locations of the $P$ most reliable sub-carriers. Using $\Sm$ we construct a $P\times N$ matrix $\Sm_P$ by pruning $\Sm$ of its zero rows. Subtracting $\Dm\lfloor\widehat{\Xfd}\rfloor$ from (\ref{eqrecsigfreq}) and using $\Sm_P$, we have
\begin{align}\label{eqformulation}
\Sm_P(\Yfd-\Dm\lfloor\widehat{\Xfd}\rfloor)&=\Sm_P\Dm(\Xfd-\lfloor\widehat{\Xfd}\rfloor)+\Sm_P\Dm\Fm\cv+\Sm_P\Zfd,
\nonumber\\
\Yfd^\prime&=\Psim\cv+\Zfd^\prime,
\end{align}
where $\Yfd^\prime=\Sm_P(\Yfd-\Dm\lfloor\widehat{\Xfd}\rfloor)$, $\Psim=\Sm_P\Dm\Fm$ and $\Zfd^\prime=\Sm_P\Zfd$. To establish the equality in aforementioned equation, we have used the fact that on RCs $\lfloor\widehat{\Xc(i)}\rfloor=\Xc(i)$, and hence $\Sm_P\Dm(\Xfd-\lfloor\widehat{\Xfd}\rfloor)=\zerov$. A typical CS problem of the form (\ref{eqformulation}), with $P$ measurements and $N$ dimensional sparse unknown ($P<<N$) can be solved using any sparse reconstruction algorithm e.g., \cite{ji2008bayesian,schniter2008fast,quadeer2012structure,blumensath2008gradient,tropp2007signal,needell2009cosamp}. However, these schemes are complex and do not utilize the clipping likelihood and undistorted phase property of the clipped signal.
\section{Proposed Clipping Reconstruction Scheme}\label{secPCRS}
From (\ref{eqclip1}) and (\ref{eqclip2}), and the discussion that followed, it is known that the clipping vector $\cv$ and the signal vector $\xv$ are anti-phased. Hence, the phase information can be deduced at the receiver from the time-domain equivalent of (\ref{eqrecsigfreqeq}), i.e., \(\hat{\xv}=\xv+\cv+\Hm^{-1}\zv\). Since $\angle{c(i)}=\angle{x(i)}-\pi~\forall~i$, only the support and the magnitudes of the active clipping elements are left unknown. Hence we can rewrite (\ref{eqformulation}) as
\begin{align}\label{eqformulationmag}
\Yfd^\prime&=\Psim\Thetam_{\cv}\cv_m+\Zfd^\prime\nonumber\\
&=\Phim\cv_m+\Zfd^\prime,
\end{align}
where $\Phim=\Psim\Thetam_{\cv}$. Here the matrix $\Thetam_{\cv}$ contains the phases of $\cv$ on its diagonal, i.e., $\Thetam_{\cv} \cong \diag(\angle{\hat{x}(0)}-\pi,\angle{\hat{x}(1)}-\pi,\cdots,\angle{\hat{x}(N-1)-\pi})$ and the vector $\cv_m$ consists of the magnitudes of the elements of $\cv$. Since the aforementioned system of equations is complex with real unknown, we can split the real and imaginary parts (designated as $\Re\{\cdot\}$ and $\Im\{\cdot\}$, respectively) to obtain a system with $2P$ equations
\begin{align}\label{eqformulationmag3}
\begin{bmatrix}\Re\{\Yfd^\prime\}\\ \Im\{\Yfd^\prime\}\end{bmatrix}&=\begin{bmatrix}\Re\{\Phim\}\\ \Im\{\Phim\}\end{bmatrix}\cv_m+\begin{bmatrix}\Re\{\Zfd^\prime\}\\ \Im\{\Zfd^\prime\}\end{bmatrix},\nonumber\\
\bar{\Yfd}&=\bar{\Phim}\cv_m+\bar{\Zfd}.
\end{align}
Henceforth, we simply use $\cv$ and not $\cv_m$ to denote the unknown signal, with the understanding that $\cv$ contains only the magnitudes and rewrite (\ref{eqformulationmag3}) as
\begin{align}\label{eqformulationmag4}
\bar\Yfd=\bar{\Phim}\cv+\bar\Zfd.
\end{align}
To solve the under-determined system in (\ref{eqformulationmag4}), we employ a Bayesian sparse reconstruction scheme. A tractable Bayesian approach (e.g., \cite{schniter2008fast}) assumes Gaussian distribution on active elements of the unknown signal. However, this is not the case here, as the nonzero elements of $\cv$ are the differences of a Rayleigh distributed elements $|x(i)|$ and a constant $\gamma$. As the unknown is clearly non-Gaussian, we pursue a Bayesian approach introduced in \cite{masood2013sparse} that does not make any assumption on the statistics of the nonzero elements of $\cv$.

Let us compute the MMSE estimate of $\cv$ given the observation $\bar{\Yfd}$ as
\begin{align}\label{eqmmseclip}
\hat\cv_{\rm mmse}\triangleq\EE[\cv|\bar{\Yfd}]=\sum_{\Sc} {\mathrm p}(\Sc|\bar{\Yc})\EE[\cv|\bar{\Yfd},\Sc],
\end{align}
where the sum is executed over all possible $2^N$ support sets $\Sc$ of $\cv$. Now assuming that the support $\Sc$ is perfectly known, (\ref{eqformulationmag4}) reduces to
\begin{equation}
\bar{\Yfd}=\bar{\Phim}_\Sc\cv_\Sc+\bar{\Zfd},
\end{equation}
where $\bar{\Phim}_\Sc$ is formed by selecting the columns of $\bar{\Phim}$ indexed by support $\Sc$. Similarly, $\cv_\Sc$ is formed by selecting entries of $\cv$ indexed by $\Sc$. Since the distribution of $\cv$ is unknown, computing $\EE[\cv|\bar{\Yfd},\Sc]$ is very difficult, if possible at all. Thus, we resort to the best linear unbiased estimate (BLUE) as an estimate of $\cv$, as given below\footnote{If $\cv$ and $\bar \Yfd$ are jointly Gaussian as is often assumed, then $\mathbb{E}[\cv|\bar\Yfd,\Sc]=\left(\Phim_\Sc\Phim_\Sc^\herm + \sigma_z^2\sigma_c^{-2}\Id\right)^{-1}\Phim_\Sc^\herm \bar\Yfd$, which applies if the statistics of $\Cfd$ and $\Zfd$ are white, Gaussian and known.}
\begin{align}
\EE[\cv|\bar{\Yfd},\Sc]\leftarrow(\bar{\Phim}_\Sc^\herm\bar{\Phim}_\Sc)^{-1}\bar{\Phim}_\Sc^\herm\bar{\Yfd}.
\end{align}

Using Bayes rule, the posterior in (\ref{eqmmseclip}) can be written as
\begin{equation}\label{eqBayesrule}
{\mathrm p}(\Sc|\bar{\Yfd})=\dfrac{{\mathrm p}(\bar{\Yfd}|\Sc){\mathrm p}(\Sc)}{{\mathrm p}(\bar{\Yfd})},
\end{equation}
where ${\mathrm p}(\bar{\Yfd})$ is common to all posteriors, and hence can be ignored. Note that Bayesian reconstruction schemes (e.g., support agnostic Bayesian matching pursuit (SABMP) \cite{masood2013sparse} and fast Bayesian matching pursuit (FBMP) \cite{schniter2008fast}) assume that the elements of the unknown are activated according to a Bernoulli distribution with success probability $\rho$. Hence, $\mathrm{p}(\Sc)$ is calculated as $\mathrm{p}(\Sc)=\mathrm{\rho}^{|\Sc|}(1-\rho)^{N-|\Sc|}$. However, for the problem at hand, it is reasonable to assume that $c(i)$ is an active element if the received sample $\hat x(i)$ is in close proximity to $\gamma$. So, instead of assigning a uniform probability $\rho$ to all samples, we assign higher probabilities to the samples that correspond to the elements of $\xv$ that are more likely to have been clipped. To do so, we define a weight vector $\wv$ with elements $w(i)=\gamma-|\hat{x}(i)|$, and assign higher probabilities to the locations where the aforementioned difference is small. One such assignment is $\rho_i=e^{-w(i)}$, where, $\rho_i$ is the probability of a clip on the $i{\rm th}$ element ($\rho_i$'s are normalized to have $\max(\rho_i)=1$). This gives us
\begin{align}\label{eqBayesruleportion1}
\mathrm{p}(\Sc)=\prod_{i\in\Sc} \rho_i \prod_{k\in\bar{\Sc}} (1-\rho_k),
\end{align}
where $\Sc\cap\bar{\Sc}=\varnothing$ and $\Sc\cup\bar{\Sc}=\{1,2,\cdots,N\}$.

We are left with the calculation of $\mathrm{p}(\bar{\Yfd}|\Sc)$, which is difficult owing to the non-Gaussian nature of $\cv_\Sc$. To get around this, we note that $\bar{\Yfd}$ is formed by a vector in the subspace spanned by the columns of $\bar{\Phim}_\Sc$ plus a Gaussian noise vector $\bar{\Zfd}$. This motivates us to eliminate the non-Gaussian component by projecting $\bar{\Yfd}$ onto the orthogonal complement space of $\bar{\Phim}_\Sc$. This is done by pre-multiplying $\bar{\Yfd}$ by a projection matrix $\Pm_\Sc^\perp$ defined as
\begin{align}
\Pm_\Sc^\perp=\Id-\Pm_\Sc=\Id-\bar{\Phim}_\Sc\left(\bar{\Phim}_\Sc^\herm \bar{\Phim}_\Sc\right)^{-1}\bar{\Phim}_\Sc^\herm.\nonumber
\end{align}
This leaves us with $\Pm_\Sc^\perp\bar{\Yfd}=\Pm_\Sc^\perp\bar{\Zfd}$, which is Gaussian with zero mean and covariance
\begin{align}
\Km&=\EE[(\Pm_\Sc^\perp\bar{\Zfd})(\Pm_\Sc^\perp\bar{\Zfd})^\herm]\nonumber\\
&=\Pm_\Sc^\perp\EE[\bar{\Zfd}\bar{\Zfd}^\herm]{\Pm_\Sc^\perp}^\herm=\Pm_\Sc^\perp\sigma_z^2{\Pm_\Sc^\perp}^\herm\nonumber\\
&=\sigma_z^2\Pm_\Sc^\perp.
\end{align}
Thus, we have,
\begin{equation}
\mathrm{p}(\bar{\Yfd}|\Sc)=\dfrac{1}{\sqrt{(2\pi\sigma_z^2)^{2P}}}\exp\left(-\dfrac{1}{2}(\Pm_\Sc^\perp\bar{\Yfd})^\herm\Km^{-1}(\Pm_\Sc^\perp\bar{\Yfd})\right).
\end{equation}
Simplifying and dropping the pre-exponential factor yields,
\begin{equation}\label{eqBayesruleportion2}
\mathrm{p}(\bar{\Yfd}|\Sc)\simeq\exp\left(-\dfrac{1}{2\sigma^2_z}\|\Pm_\Sc^\perp\bar{\Yfd}\|^2\right).
\end{equation}

Substituting (\ref{eqBayesruleportion1}) and (\ref{eqBayesruleportion2}) in (\ref{eqBayesrule}) gives the expression for posterior probability, which is then used to compute the sum in (\ref{eqmmseclip}). However, this computation is challenging as the number of support sets is large (typical values of $N$ in OFDM are $256$ and $512$). The computational burden can be reduced with a slight compromise on the performance, if this sum is computed only on the support sets corresponding to the significant posteriors $\Sc_d$ (see \cite{masood2013sparse} for details). Thus, we can write the approximated MMSE estimate of $\cv$ as
\begin{align}
\hat\cv_{\rm ammse}\triangleq\EE[\cv|\bar{\Yfd}]=\sum_{\Sc_d} \mathrm{p}(\Sc|\bar{\Yc})\EE[\cv|\bar{\Yfd},\Sc].
\end{align}
Now, we pursue a greedy approach \cite{schniter2008fast,masood2013sparse} to find a subset of the dominant support $\Sc_d$. Note that though this approach of sparse signal reconstruction was presented in \cite{masood2013sparse}, the proposed clipping recovery scheme has two differentiating characteristics. First is the use of the weighted $\mathrm{p}(\Sc)$ in (\ref{eqBayesruleportion1}), which helps to find the dominant support much faster than the un-weighted case. Second is the phase augmentation, which results in improved reconstruction accuracy.

The Bayesian reconstruction approach discussed above relies on the \emph{a priori} information about the sparsity rate $\rho$, the noise variance $\sigma_z^2$ and the clipping threshold $\gamma$ to reconstruct the vector $\cv$. The threshold $\gamma$ can be communicated to the receiver during the signalling period, $\rho$ can be obtained from previously accumulated data and any SNR estimation scheme can be used to find $\sigma_z^2$. Nonetheless, if accurate estimates of these quantities are not available, the proposed scheme can bootstrap itself and estimate these parameters from the data. Specifically, in the absence of accurate estimates, we start with initial rough estimates of the parameters and obtain $\hat{\cv}$. The estimate of $\cv$ is then used to refine the parameters $\hat{\sigma}_z^2$ and $\hat{\rho}$, and these parameters are now used to obtain an improved estimate of $\cv$. This procedure can continue iteratively, until a predetermined criterion is satisfied. The computational complexity of the proposed reconstruction scheme is of the order $\Oc(E_{\rm max}\rho PN^2)$, if an $N$-dimensional signal with $\rho N$ non-zero elements is estimated using $P$ measurements and the parameter refinement is performed $E_{\rm max}$ times \cite{masood2013sparse}. As the proposed scheme uses weighting and phase augmentation we term it weighted and phase augmented (WPA)-SABMP. An algorithmic description of the WPA-SABMP reconstruction scheme is provided in Table~\ref{tablesummary}.
\begin{table}[t!]\centering
\caption{Summary of the proposed WPA-SABMP scheme}
\label{tablesummary}
\normalsize
\begin{tabular}{c}
\toprule
\begin{minipage}{0.45\textwidth}
    \vskip 4pt
    \begin{enumerate}
   \item \emph{Equalize:} $\hat{\xv}=\Fm^\herm\Dm^{-1}\Fm\yv$
   \item \emph{Estimate clipping level:} $\hat{\gamma}=\max(\hat{\xv})$
   \item \emph{Calculate the weight:} $\wv=\hat{\gamma}-|\hat{\xv}|$
   \item \emph{Find Reliable Carriers:} Calculate $\mathfrak{R}$ and choose $P$ carriers with the highest reliability.
   \item \emph{Estimate the sparsity rate:} $\hat{\rho}(0)=\sf{Q}\left(\dfrac{\hat{\gamma}-\mu}{\sigma}\right)$, an initial estimate, where $\mu$ and $\sigma$ are the mean and standard deviation of $\hat\xv$, respectively.
   \item For $t=1,2,\cdots$, \emph{repeat}
   \begin{enumerate}
   \item $\hat\rho(t)_i=\hat{\rho}(t)e^{-w(i)},~i=1,2,\cdots,N$
   \item \emph{Compute:} $\hat{\cv}_{\rm ammse}$ and $\hat{\rho}(t)$ using the technique discussed in \cite{masood2013sparse} \\\emph{until} $\left(\dfrac{|\hat{\rho}(t)-\hat{\rho}(t-1)|}{\hat{\rho}(t-1)}<0.02\right)$
   \item \emph{Phase augment:} $\hat{\cv}=\Thetam_c|\hat{\cv}_{\rm ammse}|$
   \item \emph{Remove distortion:} $\hat{\xv}(t+1)=\hat{\xv}(t)-\hat{\cv}$
   \end{enumerate}
   \end{enumerate}
   \vskip 4pt
 \end{minipage}
 \\
\bottomrule
\end{tabular}
\end{table}
\subsection{Simulation Results}
The SABMP algorithm was proposed in \cite{masood2013sparse} and was shown to outperform other Bayesian and $\ell_1$-based sparse recovery algorithms. Hence, in this work we compare the proposed WPA-SABMP scheme with SABMP \cite{masood2013sparse}, the phase augmented version of FBMP i.e., PA-FBMP and the weighted and phase augmented-LASSO (WPAL) \cite{al2012peak}. As a benchmark, we use the oracle-least squares (LS) solution (i.e., the case when the support is perfectly known and the LS solution is calculated on the known support). In all simulations it is assumed that the statistics (i.e., the mean and the variance) of the clipping signal are not known. These schemes are compared for their BER performance and practical complexity. The practical complexity is calculated as the average runtime for signal recovery and is presented by subgraphs within the main figures (the independent axes of the subgraph and the corresponding main figure are always identical).

An OFDM system with $512$ sub-carriers is simulated. The $64$-QAM alphabet is used for modulation and the data is passed through a channel with $10$ i.i.d. taps of unit variance. All simulation results are averaged over $5000$ bit errors unless otherwise noted.
\subsubsection{Experiment 1}
\begin{figure}[h!]\centering
\includegraphics[width=0.45\textwidth]{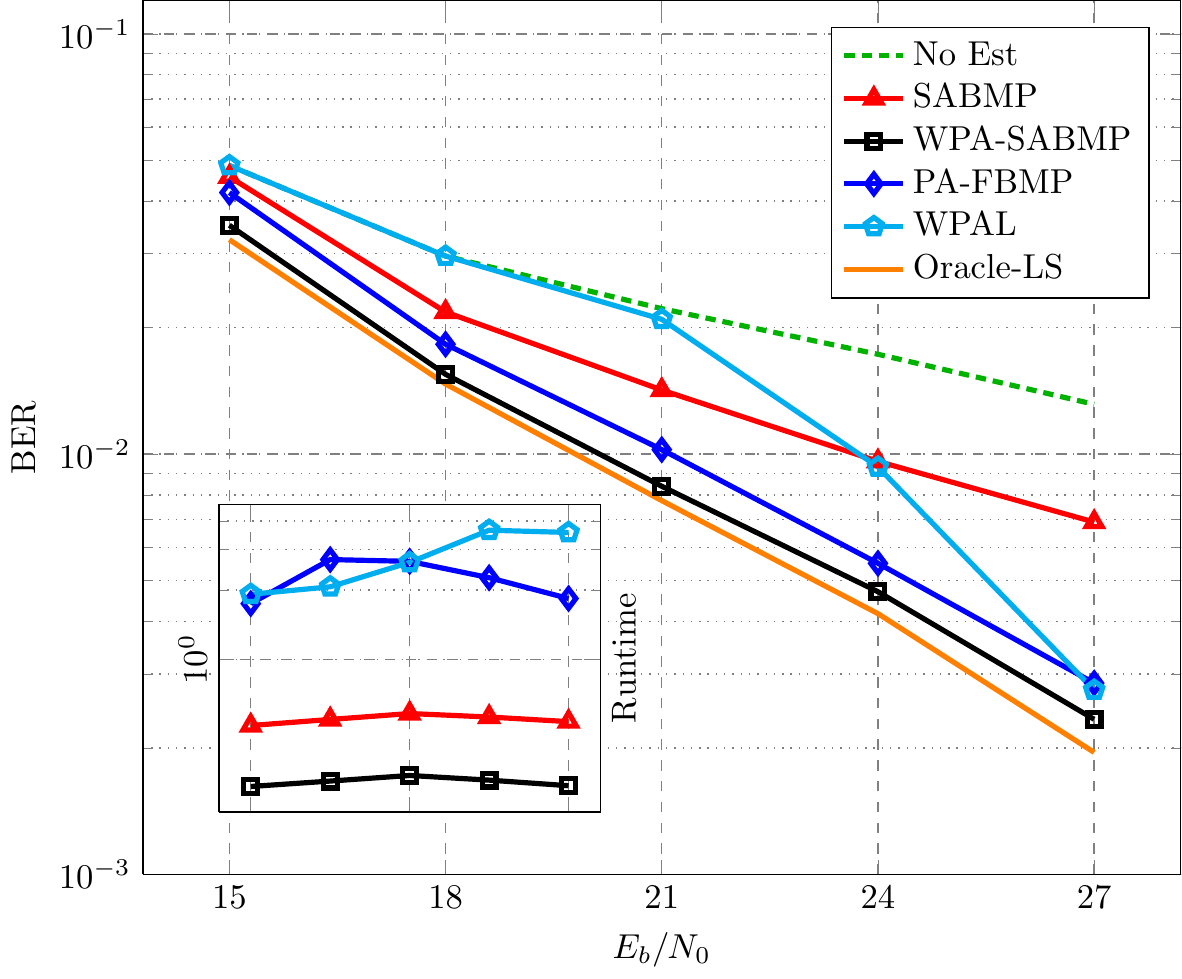}
\caption{BER versus $E_b/N_0$ ($\mathrm{CR}=1.61$,$P=128$).}
\label{FigExp1_SNR}
\end{figure}

In this experiment, sparse reconstruction schemes are tested for their BER performance. The $\mathrm{CR}$ is kept fixed at $\mathrm{CR}=1.61$ and the number of RCs is set to $P=128$. It can be observed from the results in Fig.~\ref{FigExp1_SNR} that the proposed scheme provides significant gain over existing reconstruction schemes and attains a BER very close to the oracle-LS. Further, it can be noticed from the subgraph that among the compared schemes, WPA-SABMP is the least complex clipping reconstruction scheme.

\subsubsection{Experiment 2}
\begin{figure}[h!]\centering
\includegraphics[width=0.45\textwidth]{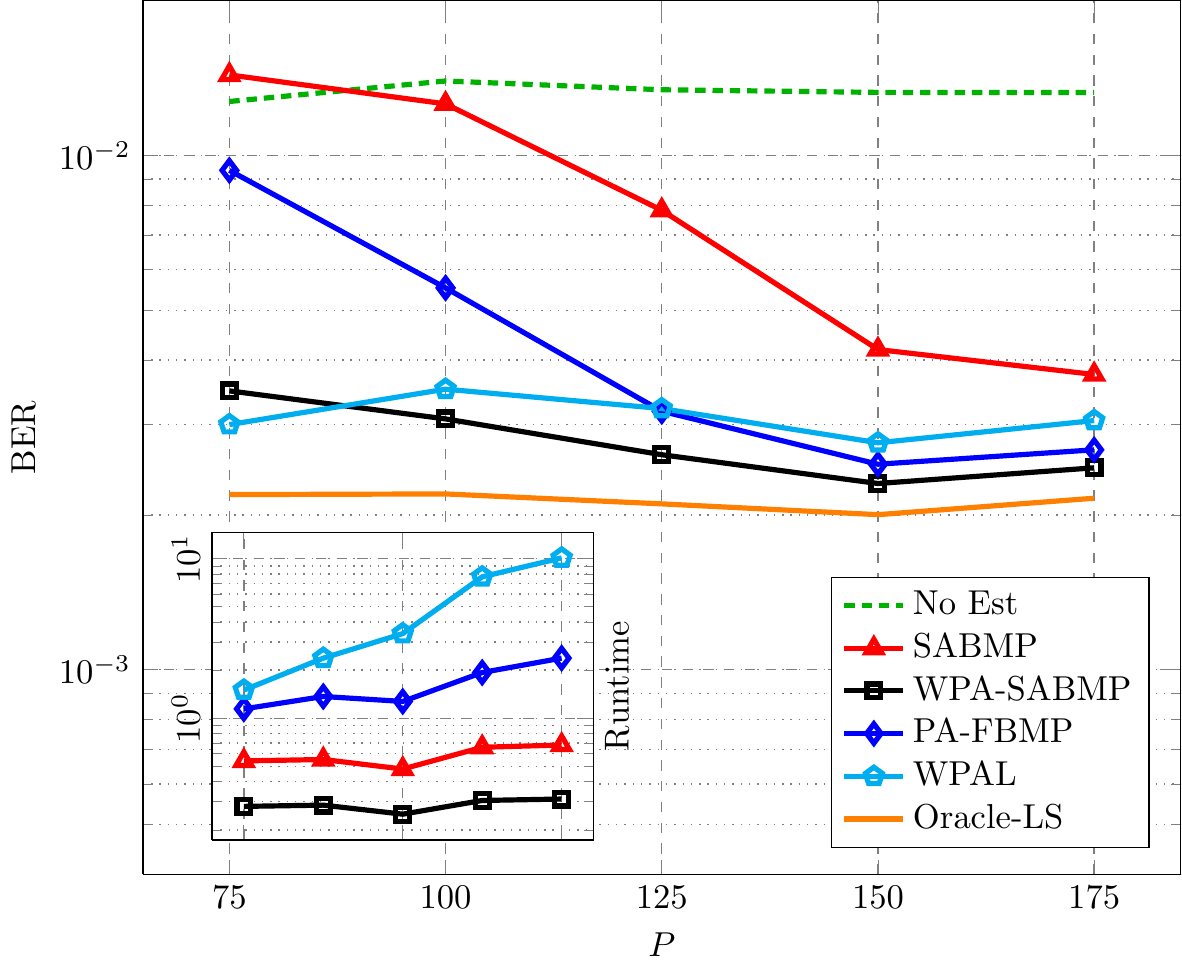}
\caption{BER versus $P$ ($\mathrm{CR}=1.61$,$E_b/N_0=27$dB).}
\label{FigExp1_RelCar}
\end{figure}

In this experiment, $E_b/N_0$ is kept fixed at $27$ dB and the number of RCs $P$ used for reconstruction is varied from $75$ to $175$. Observe (from Fig.~\ref{FigExp1_RelCar}) that if $P$ is reduced, the reconstruction accuracy of SABMP and PA-FBMP is reduced, however, WPA-SABMP and WPAL show robustness against reduced $P$. Though WPAL has good reconstruction accuracy in the range of interest, it is the most complex algorithm among the compared schemes. Further, this complexity is elevated with increasing $P$. The time graph also shows that the WPA-SABMP has least complexity and it varies only slightly with $P$.
\subsubsection{Experiment 3}
\begin{figure}\centering
\includegraphics[width=0.45\textwidth]{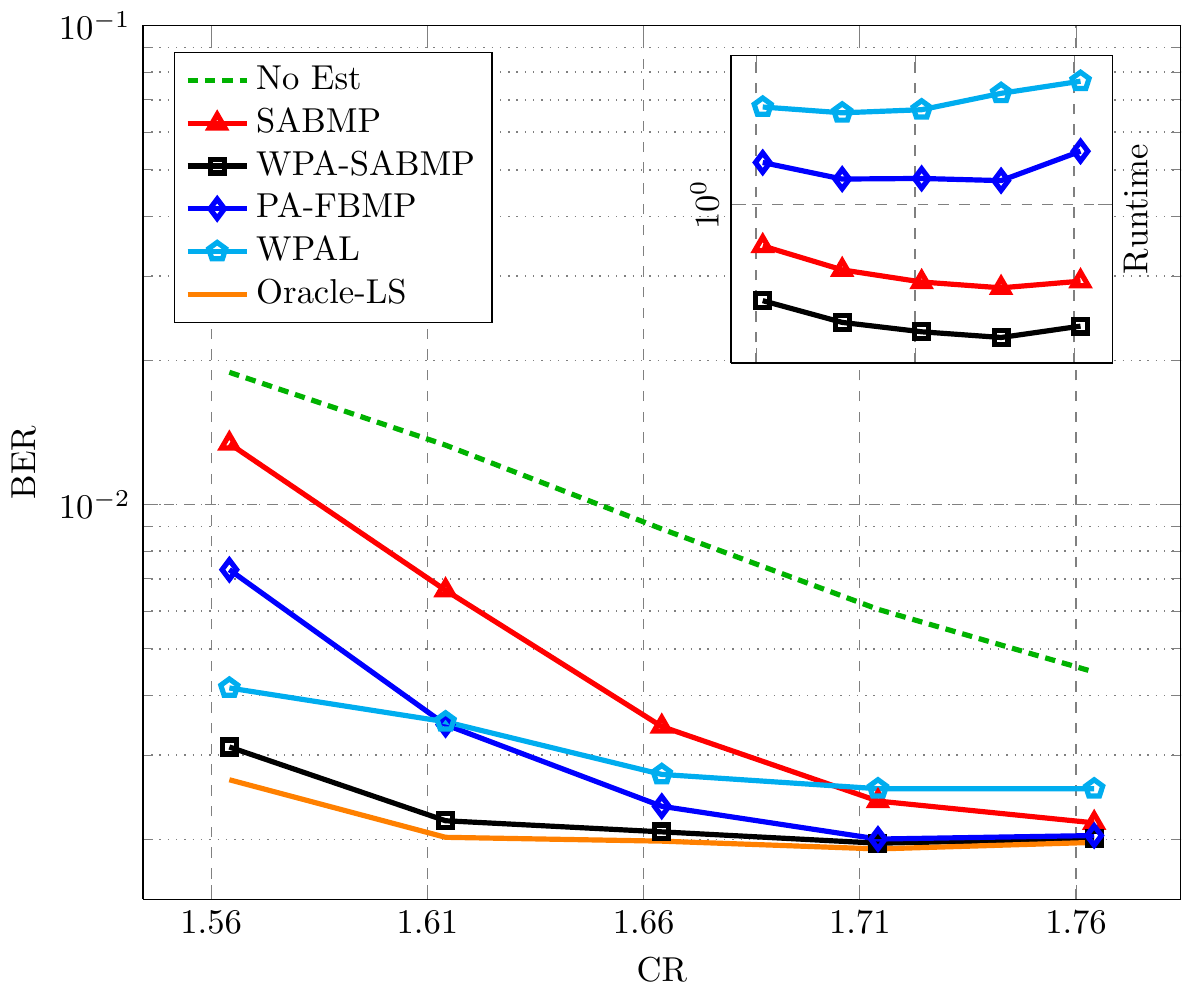}
\caption{BER versus $\mathrm{CR}$ ($P=128$,$E_b/N_0=27$dB).}
\label{FigExp1_CR}
\end{figure}

In this experiment, the performance of the proposed scheme is tested vs the $\mathrm{CR}$ while keeping $E_b/N_0$ and $P$ fixed. It is natural that the performance of the reconstruction schemes improves as the clipping is reduced (i.e., for higher $\mathrm{CR}$ values). However, as shown in Fig.~\ref{FigExp1_CR}, the proposed WPA-SABMP scheme performs better than SABMP and PA-SABMP for all $\mathrm{CR}$ values and better than WPAL for most $\mathrm{CR}$ values. Further, observe that the WPA-SABMP scheme recovers the clipping in a small time irrespective of the $\mathrm{CR}$.
\subsubsection{Experiment 4}
\begin{figure}\centering
\includegraphics[width=0.45\textwidth]{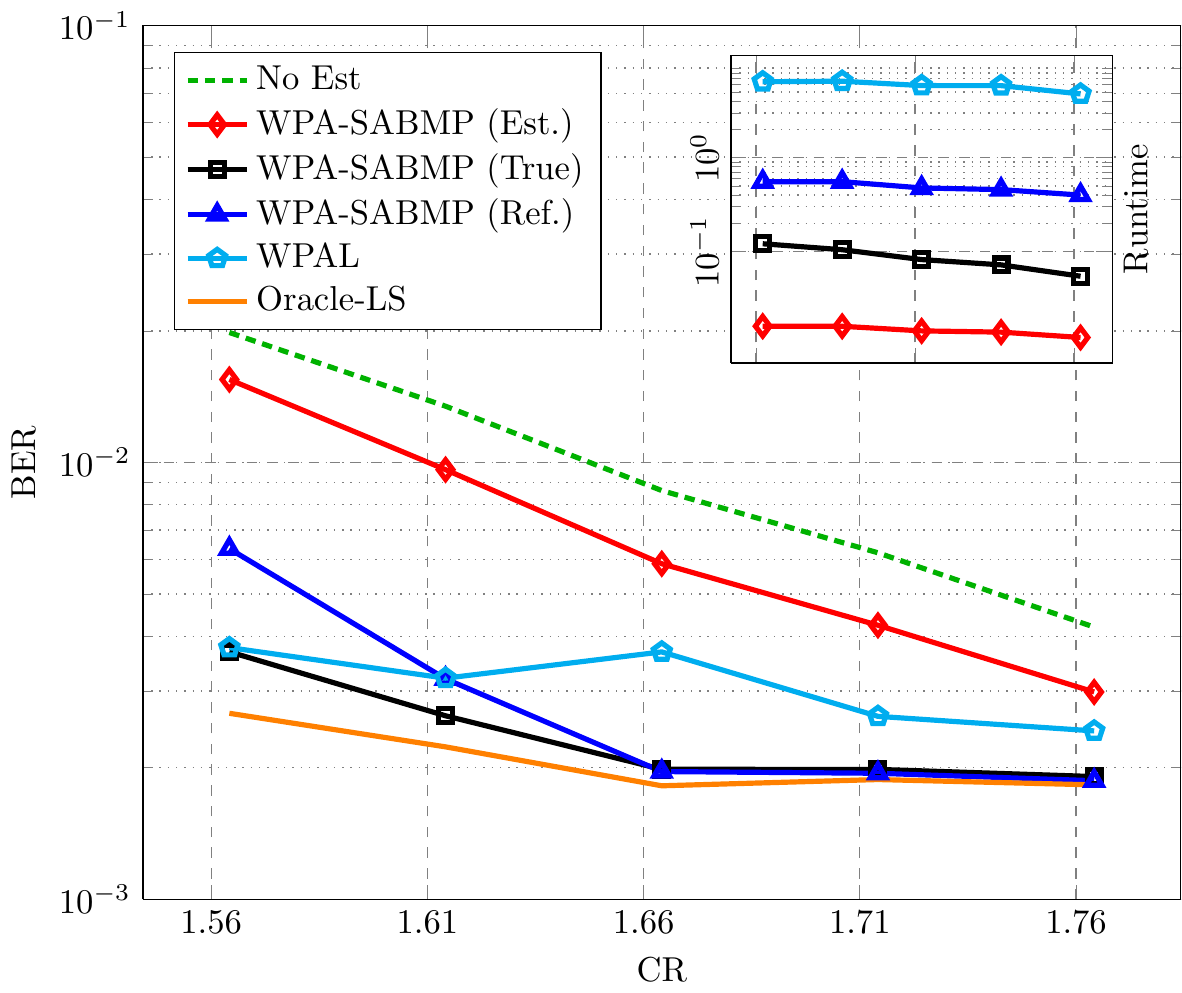}
\caption{BER versus $\mathrm{CR}$ ($P=128$, $E_b/N_0=27$dB, $\rho_{\rm init}=0.01\rho_{\rm true}$ and $\sigma_{z\_\rm init}^2=0.01\sigma_{z\_\rm true}^2$).}
\label{FigExp1_CR_Ref}
\end{figure}

In this experiment, we compare the performance of the proposed scheme in the absence of accurate estimates of the signal statistics (i.e., the threshold $\gamma$, the noise variance $\sigma_z^2$ and the sparsity rate $\rho$ are not known). The results of this experiment are depicted in Fig.~\ref{FigExp1_CR_Ref}. The WPA-SABMP (True) scheme in Fig.~\ref{FigExp1_CR_Ref} represents the case when the actual estimates are available, WPA-SABMP (Est.) represents the case where the actual estimates are not available but no refinement is performed and WPA-SABMP (Ref.) represents the case where the proposed scheme is run $E_{\rm max}=5$ times for refinement of the initial estimates. The proposed refinement-based scheme is compared with WPAL as it does not require any signal statistics. The initial estimates of the signal sparsity and noise variance are $\rho_{\rm init}=0.01\rho_{\rm true}$ and $\sigma_{z\_\rm init}^2=0.01\sigma_{z\_\rm true}^2$. It is observed that using the refinement procedure, even in the absence of accurate statistics, performance very close to the oracle-LS can be obtained. However, as the refinement procedure runs  $E_{\rm max}$ times, it requires more execution time than its non-refined counterpart.
\section{Clipping Reconstruction for SIMO Systems}\label{secSIMOS}
Let us consider an OFDM communication system equipped with $L$ receiver antennas. At the receiver we have $L$ independent copies of the transmitted signal. All diversity branches contain the same distortion signal $\cv$, convolved with the channel impulse response $\hv_l$ of the $l{\rm th}$ branch. For acceptable performance of the communication system, the distortion needs to be eliminated from all diversity branches before the signals are combined to obtain an estimate of the transmitted signal. The distortion-free independent versions of the received signal can be combined using any of the well-known diversity combining methods (e.g., equal gain combining (EGC), selection combining (SC) and maximal ratio combining (MRC) \cite{boutros1998signal}) to obtain an estimate of the transmitted signal.

To pursue the reconstruction of $\cv$ using the scheme proposed in Section \ref{secPCRS}, a system of equations of the form (\ref{eqformulationmag4}) is formulated for each diversity branch. In general, for the $l{\rm th}$ branch we have
\begin{equation}
\bar\Yfd_{l}=\bar{\Phim}_{l}\cv+\bar\Zfd_{l},
\label{eqmulante1}
\end{equation}
where $\bar\Yfd_l$ is the measurement vector associated with the $l{\rm th}$ diversity branch of the system (similar definitions apply to $\bar\Phi_l$ and $\bar\Zfd_l$). Note that $\cv$ is free of subscript $l$, as it is same \emph{for all diversity branches}.

\begin{figure}\centering
\includegraphics[width=0.45\textwidth]{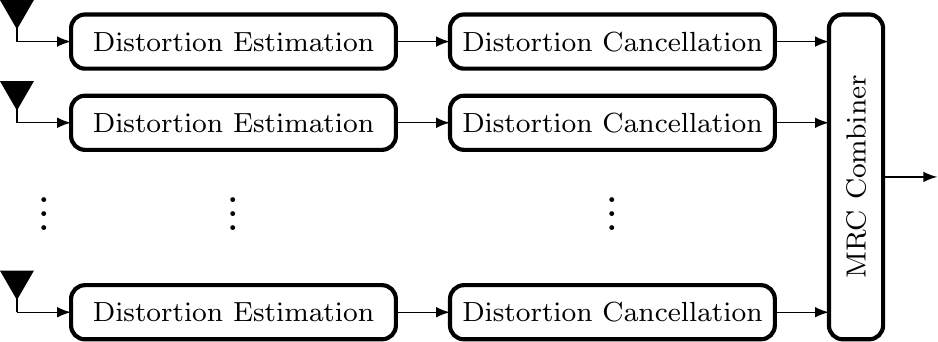}
\caption{Individual reconstruction per diversity branch.}
\label{figindrecons}
\end{figure}

One rather obvious approach towards estimation of $\cv$ given $L$ systems of the form (\ref{eqmulante1}) is \emph{individual reconstruction} per diversity branch as shown in Fig.~\ref{figindrecons}. Once the estimates of the clipping distortion are available, they are subtracted from the respective branches to obtain the distortion-free versions $\check{\Yfd}_l=\Yfd_l-\hat\cv$ of the transmitted signal corresponding to each branch. These signals are then combined using MRC to obtain $\widehat{\Xfd}$ using the following expression
\begin{equation}\label{eqMRC}
\widehat{\Xfd}=\sum_{l=1}^{L}\Dm_l^\herm\check{\Yfd}_l,
\end{equation}
where $\Dm_l$ is the diagonal frequency response matrix corresponding to the $l{\rm th}$ branch.
An alternative and more effective approach is to utilize the fact that the clipping signal is same over all diversity branches. In this relation, the $L$ systems of linear equations (\ref{eqmulante1}) can be concatenated and setup in the following form:
\begin{equation}
\begin{bmatrix}
\bar\Yfd_{1}\\
\bar\Yfd_{2}\\
\vdots\\
\bar\Yfd_{L}\\
\end{bmatrix}
=
\begin{bmatrix}
\bar\Phim_{1}\\
\bar\Phim_{2}\\
\vdots\\
\bar\Phim_{L}\\
\end{bmatrix}
\cv+
\begin{bmatrix}
\bar\Zfd_{1}\\
\bar\Zfd_{2}\\
\vdots\\
\bar\Zfd_{L}\\
\end{bmatrix},
\end{equation}
which can be written more compactly as
\begin{equation}
\overrightarrow{\Yfd}=\overrightarrow{\Phim}\cv+\overrightarrow{\Zfd}.
\end{equation}

\begin{figure}\centering
\includegraphics[width=0.35\textwidth]{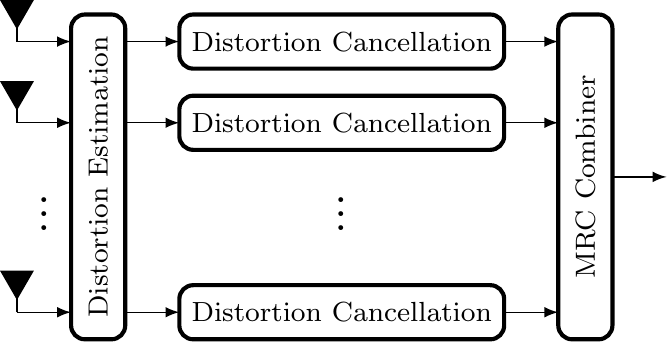}
\caption{Joint reconstruction over all diversity branches.}
\label{figjntrecons}
\end{figure}

It is evident that with $2P$ measurements per diversity branch, a total of  $2LP$ measurements are now available to reconstruct the sparse unknown (see Fig.~\ref{figjntrecons}). Once $\hat{\cv}$ is obtained as done in Section~\ref{secPCRS} for the single antenna case, the subsequent distortion removal and MRC combining is identical to the case of individual reconstruction.
\subsection{Simulation Results}
\begin{figure}\centering
\includegraphics[width=0.45\textwidth]{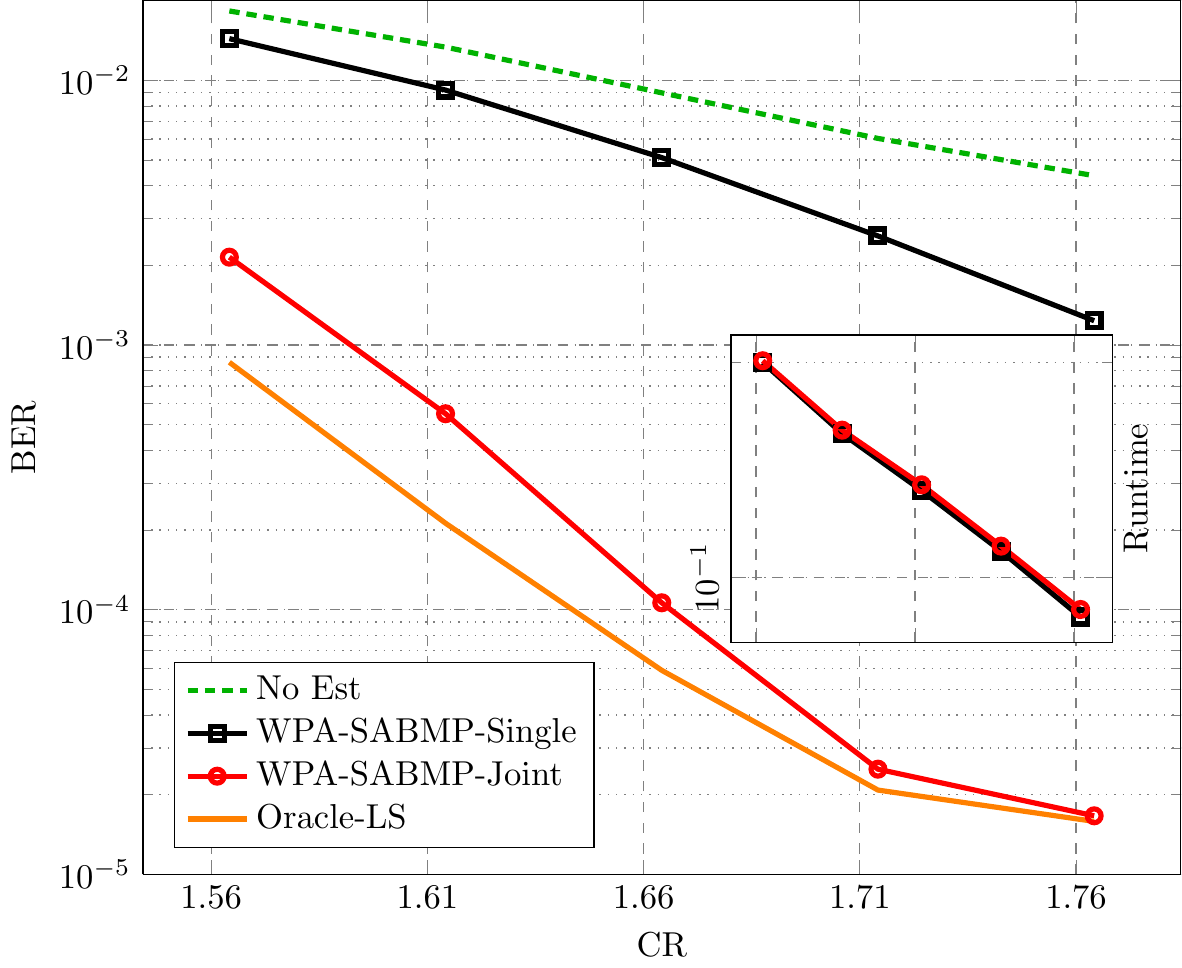}
\caption{BER versus $\mathrm{CR}$ for SIMO-OFDM Communication Systems ($P=77$,$E_b/N_0=27$dB).}
\label{FigExp2_CR}
\end{figure}

In this experiment, the performance of the proposed joint reconstruction scheme is compared with individual reconstruction for two antennas at the receiver i.e., $L=2$. The $\mathrm{CR}$ is varied while $E_b/N_0$ and $P$ are kept fixed. The simulation is averaged over $500$ bit errors. The results in Fig.~\ref{FigExp2_CR} show that the joint reconstruction scheme achieves an error rate much lower than individual reconstruction. Further, to compare the computational complexity, we note that individual reconstruction can be performed in parallel, therefore we consider the time required for signal reconstruction in one branch only. It is observed from the subgraph that the average time taken by the joint and individual reconstruction is almost the same.
\section{Multi-user Communication}\label{secMUComm}
In multi-user OFDM systems i.e., OFDMA, each user is assigned a subset of sub-carriers, and each sub-carrier is assigned exclusively to one user \cite{kivanc2003computationally}. The time-domain signal resulting from an IDFT on each user is clipped for PAPR reduction. Clipping multiple users simultaneously complicates the estimation process at the receiver. This is because the distortion from each user is spread over all sub-carriers and hence overlap. The frequency-domain overlap of distortion renders many of the assumptions made in the single-user scenario invalid. To be specific, weighting and phase augmentation cannot be applied in multi-user clipping estimation directly. Further, as the data on each sub-carrier is corrupted by clipping distortion from all users (and additive noise), the perturbations are generally strong enough to take the data out of the corresponding decision regions and hence the RC method is inapplicable. Hence, in multi-user clipping estimation, we resort to data-free pilot tones for measuring the clipping distortion.

Let us commence the formulation of a multi-user clipping estimation strategy by generalizing the data model presented in Section~\ref{secDatMod} for OFDMA systems. In this work, we consider the two-user case for clarity of exposition; however, the proposed scheme is easily extendable to the general $U$ user case. In the uplink of an OFDMA system, the total number of available sub-carriers $N$ is divided between the two subscribers and each user will be allocated $K=N/2$ sub-carriers for data transmission. The sub-carriers can be allocated adjacently (sub-carriers $(u-1)N/K$ to $uN/K-1$ are reserved for the $u{\rm th}$ user) or in an interleaved manner (user $u$ is allocated sub-carriers $u+dK-(K+1),~~ d\in\{1,2,\cdots,N/K\}$). In this work, we focus solely on interleaved carrier allocation. In the context of a complete OFDMA symbol, the frequency-domain signal corresponding to the first user can be written as
\begin{align}\centering
\Xfd^1=&[\Xc^1(0), 0, \Xc^1(1), 0, \cdots,\Xc^1(N/2-1),0]^\transp,\nonumber
\end{align}
and the signal corresponding to the second user is given by
\begin{align}\centering
\Xfd^2=&[0, \Xc^2(0), 0, \Xc^2(1), 0, \cdots,\Xc^2(N/2-1)]^\transp.\nonumber
\end{align}
The time-domain signal for the $u{\rm th}$ user (i.e., $\xv^u$) is obtained by taking the IDFT of $\Xfd^u$. To reduce the PAPR, the signals $\xv^{u}$ are clipped as given by (\ref{eqclip1}); at the receiver we have \(\yv=\Hm^1(\xv^1+\cv^1) + \Hm^2(\xv^2+\cv^2) + \zv\). The frequency-domain received signal can be obtained by the DFT operation as
\begin{equation}
\label{recfredom}
\Yfd=\Dm^1\Xfd^1+\Dm^1\Cfd^1+\Dm^2\Xfd^2+\Dm^2\Cfd^2+\Zfd.
\end{equation}

Note that, although the channel frequency responses \(\Dm^u\) are diagonal matrices of size $N\times N$ and hence are overlapping, the matrix $\Dm$ comprises only the portions of $\Dm^u$, belonging to the $u{\rm th}$ user band, which is denoted by $\underline{\Dm}^u$. Hence, we can write
\begin{equation}
\Yfd=\Dm\Xfd+\Dm^1\Cfd^1+\Dm^2\Cfd^2+ \Zfd,
\label{recfredomsim}
\end{equation}
where $\Xfd=\Xfd^1+\Xfd^2$. In the absence of distortion (i.e., when $\Dm^1\Cfd^1=\Dm^2\Cfd^2=\zerov$), the receiver could easily separate the users (as they occupy different carriers) and equalize the users' channels (as in (\ref{eqrecsigfreqeq})) to recover the transmitted data. Mathematically, we can write $\underline{\Yfd}^u=\underline{\Dm}^u\underline{\Xfd}^u+\underline{\Zfd}^u$, where $\underline{\Yfd}^u$ is the portion of $\Yfd$ confined to the carriers of the $u{\rm th}$ user (a similar definition applies to $\underline{\Dm}^u$,$\underline{\Xfd}^u$ and $\underline{\Zfd}^u$). Upon equalizing, we obtain
\begin{equation}\label{eq:MLdec}
\widehat{\underline{\Xfd}^u}=(\underline{\Dm}^u)^{-1}\underline{\Yfd}^u=\underline{\Xfd}^u+(\underline{\Dm}^u)^{-1}\underline{\Zfd}^u.
\end{equation}
The noisy estimate $\widehat{\underline{\Xfd}^u}$ is then rounded to the nearest constellation point ($\lfloor\widehat{\underline{\Xfd}^u}\rfloor$). However, in the presence of distortion, clipping needs to be estimated and cancelled before the equalization step of (\ref{eq:MLdec}).

Now to demonstrate how clipping distortion can be estimated, we re-write (\ref{recfredomsim}) as
\begin{align}
\!\!\!\Yfd\!\!&=\!\!\Dm\Xfd\!+\![\Dm^1 \Dm^2]\!\begin{bmatrix}\Cfd^1\\\Cfd^2\end{bmatrix}\!+\!\Zfd\nonumber\\
&=\Dm\Xfd\!+\![\Dm^1 \Dm^2]\begin{bmatrix}\Fm&\zerov\\\zerov&\Fm\end{bmatrix}\begin{bmatrix}\cv^1\\\cv^2\end{bmatrix}\!+\!\Zfd,
\label{recfredomsim2o}
\end{align}
where we have made the substitution $\Cfd^u=\Fm\cv^u$.  Using a selection matrix $\Sm_P$ we proceed by projecting $\Yfd$ onto the subspace spanned by the reserved carriers. This yields
\begin{align}\label{eq:basicCSSys}
\Sm_P\Yfd\!\!&=\!\!\Sm_P(\Dm\Xfd\!+\![\Dm^1 \Dm^2]\begin{bmatrix}\Fm&\zerov\\\zerov&\Fm\end{bmatrix}\!\cv\!+\!\Zfd)\nonumber\\
{\text{i.e.,}}~\Yfd^{\prime}\!\!&=\!\!\Psim\cv\!+\!\Zfd^\prime.
\end{align}

The clipping $\cv$ can be recovered from the under-determined systems in (\ref{eq:basicCSSys}) by sparse signal reconstruction. However, the assumptions used for weighting and phase augmentation in earlier parts of this paper are no longer valid. Though the signal can be recovered using sparse signal recovery tools (e.g., FBMP, SABMP and $\ell_1$-optimization), in the multiuser scenario this is not really effective especially as the number of users gets larger. For example, in the two-user scenario of (\ref{recfredomsim}), the sparse vector is twice as large and could have twice the number of active elements. As such, to maintain the quality of the estimate in two-user scenario, we need to double the number of free carriers, which will reduce the throughput. Alternatively, here we get by with the estimate obtained from (\ref{eq:basicCSSys}) and once these estimates are available we proceed in a decoupled manner to improve them.

Once the clipping signals are initially reconstructed using (\ref{eq:basicCSSys}) (i.e., joint estimation), it is possible to set up two uncoupled systems of equations for user $1$ and $2$ respectively. After the isolated systems are formed, sparse clipping reconstruction can be performed for each user for enhanced recovery. Therefore, the crux of the proposed reconstruction scheme can be summarized in the following two steps: 1) Estimate $\cv=[{\cv^1}^\transp {\cv^2}^\transp]^\transp$ via joint sparse reconstruction using (\ref{eq:basicCSSys}) and 2) Decouple the two systems of linear equations corresponding to user $1$ and user $2$ and perform clipping reconstruction for each user.

To obtain the decoupled systems, we modify the approach initially proposed for channel estimation \cite{wu2005iterative} (we term this approach the contaminated pilot approach). It was noted that as the clipped signal is transmitted (transmitted pilots are also clipped) it is not optimal to use the ideal pilot sequence at the receiver as a reference for channel estimation. Instead, the clipped pilot sequence was first estimated at the receiver and then used for enhanced channel estimation. As the clipped pilots are used in \cite{wu2005iterative} instead of clean pilot signals, we call this scheme the contaminated pilot approach. In this work, we use the idea of reconstructing the clipped version of the transmitted signal at the receiver to form the decoupled systems. To do that, the initial estimate of $\cv$ obtained using (\ref{eq:basicCSSys}) is subtracted from (\ref{recfredomsim}) to get
\begin{align}
\Yfd_{\rm cs}=\Yfd-[\Dm^1 \Dm^2]\begin{bmatrix}\Fm&\zerov\\\zerov&\Fm\end{bmatrix}\hat\cv=\Dm\Xfd+\Zfd^\prime.
\end{align}

We proceed by extracting the carriers allocated to user $u$ and get $\underline{\Yfd}^u_{\rm cs}$, which is then equalized using (\ref{eq:MLdec}) to obtain $\widehat{\underline{\Xfd}^u}=(\underline{\Dm}^u)^{-1}\underline{\Yfd}^u_{\rm cs}$. Now, we estimate the transmitted frequency-domain signal by making the ML decisions  $\lfloor\widehat{\underline{\Xfd}^u}\rfloor$. The time-domain signal is obtained by IDFT as $\widehat{\xv^u}=\Fm^\herm\lfloor\widehat{\underline{\Xfd}^u}\rfloor$. This time-domain signal is then clipped using (\ref{eqclip1}) to get $\widehat{\xv_p^u}$. Now the difference between the clipped and un-clipped versions of $\widehat{\xv}^u$ i.e., $\widehat{\cv}^u=(\widehat{\xv}_p^u-\widehat{\xv}^u)$ is entrusted as the improved estimate of the clipping distortion and is subtracted from (\ref{recfredomsim}) to form the decoupled systems. The stepwise procedure for formulation of the decoupled system is outlined below:

\begin{enumerate}
\item Perform joint sparse signal reconstruction based on (\ref{eq:basicCSSys}).
\item Subtract the estimated distortion $\hat\cv$ from (\ref{recfredomsim}) to obtain\\
$\Yfd_{\rm cs}=\Yfd-[\Dm^1 \Dm^2]\begin{bmatrix}\Fm&\zerov\\\zerov&\Fm\end{bmatrix}\hat\cv=\Dm\Xfd+\Zfd\nonumber$
\item Get $\underline{\Yfd}_{\rm cs}^u=\underline{\Dm}^u\Xfd^u\!\!+\!\!\underline{\Zfd}^u\nonumber$ by extracting user $u$'s sub-carriers.
\item Equalize $\underline{\Yfd}^u_{\rm cs}$ using (\ref{eq:MLdec}) and obtain ($\lfloor\widehat{\underline{\Xfd}^u}\rfloor$).
\item Using pilots and $\lfloor\widehat{\underline{\Xfd}^u}\rfloor$, form a time-domain signal $\widehat{\xv^u}$.
\item Obtain $\widehat{\xv_p^u}$ from $\widehat{\xv^u}$ using (\ref{eqclip1}) and obtain $\widehat{\Cfd^u}=\widehat{\Xfd}_p^u-\widehat{\Xfd^u}$.
\item Obtain $\Yfd^{\bar u}\!\!=\!\!\Yfd\!-\!\Dm^u\widehat{\Cfd^u}\!\!=\!\!\Dm\Xfd\!+\!\Dm^u\Cfd^u\!+\!\Zfd,~\bar u\neq u$ based on (\ref{recfredomsim}).
\end{enumerate}

Note that $\Yfd^{\bar u}$ is decoupled from user $u$'s clipping. Now, with this decoupled system for user $\bar u$, we can extract the sub-carriers allocated to user $\bar u$ to form $\underline{\Yfd}^{\bar u}=\underline{\Dm}^{\bar u}\underline{\Xfd}^{\bar u}+\underline{\Dm}^{\bar u}\Cfd^{\bar u}+\underline{\Zfd}^{\bar u}$ and reconstruct $\cv^{\bar u}$ using sparse recovery.
\subsection{Simulation Results}
\begin{figure}\centering
\includegraphics[width=0.45\textwidth]{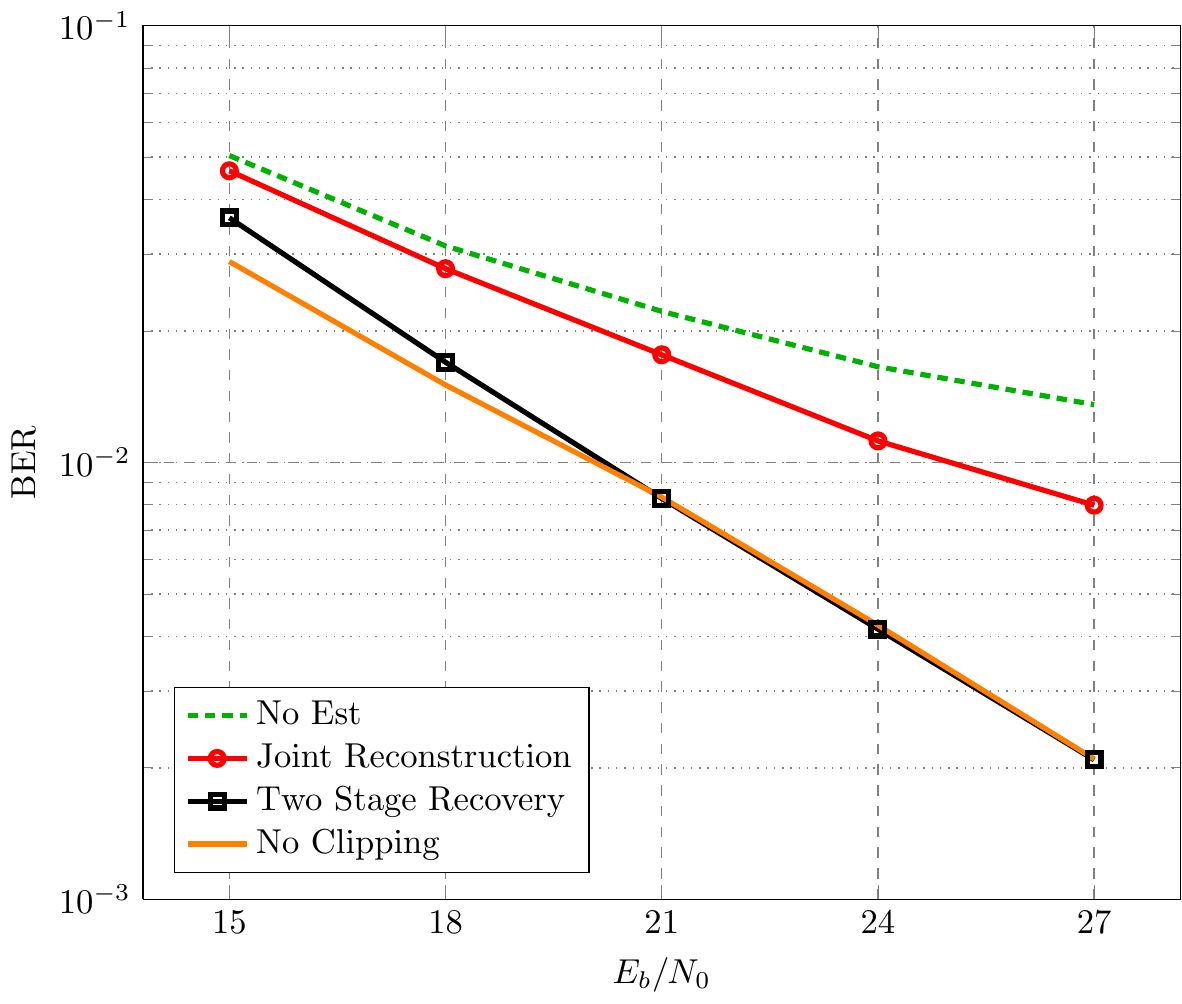}
\caption{BER versus $E_b/N_0$ for Multi-user clipping recovery scheme ($\mathrm{CR}=1.61$,$P_u=75$).}
\label{FigMU}
\end{figure}

The OFDMA system with two users is simulated using $512$ sub-carriers and $64$-QAM modulation. Each user is assigned a total of $256$ sub-carriers in an interleaved fashion. The number of reserved tones used for CS measurements is $P_u=75~\text{for}~u=1,2$. The threshold for both users is chosen such that $\mathrm{CR}=1.61$. For sparse signal reconstruction FBMP \cite{schniter2008fast} is used and the results are presented (in Fig.~\ref{FigMU}) that compare the proposed (two-stage recovery) scheme with the joint estimation scheme. It can be seen that joint reconstruction of the clipping distortion gives very little gain in BER. However, the proposed decoupling-based two-stage multi-user clipping reconstruction scheme significantly improves the BER and achieves the no clipping rate for high $E_b/N_0$.
\section{Channel Estimation in the Presence of Clipping}\label{secCEPC}
Clipping the transmission signal results in pilot contamination, hence the MMSE estimation based on these pilot signals is not optimal. In this section, we discuss the channel estimation problem for clipped OFDM and present data-aided CIR estimation strategies.

The received OFDM signal is given in (\ref{eqrecsigfreq}) and can be written as
\begin{align}\label{eqrecsigfreq2}
\Yfd=\Dm\Xfd+\Dm\Cfd+\Zfd=\Dm\Xfd+\Zfd^\backprime,
\end{align}
where $\Zfd^\backprime=\Dm\Cfd+\Zfd$ is the combined AWGN noise and clipping distortion. Let us define $\Dfd\triangleq\diag(\Dm)=\sqrt{N}\bar\Fm\hv$ (where $\bar\Fm$ represents the $N\times N_c$ partial DFT matrix obtained by pruning $\Fm$ of its last $N-N_c$ columns). Now note that $\Dm\Xfd$ is a product of a diagonal matrix and a column vector, and hence we can exchange the roles of $\Dm$ and $\Xfd$ by rewriting (\ref{eqrecsigfreq2}) as
\begin{align}\label{eqrecsigfreq3}
\Yfd&=\diag(\Xfd)\diag(\Dm)+\Zfd^\prime=\diag(\Xfd)\Dfd+\Zfd^\prime\nonumber\\
&=\sqrt{N}\diag(\Xfd)\bar\Fm\hv+\Zfd^\prime=\Xm\hv+\Zfd^\prime,
\end{align}
where $\Xm\triangleq\sqrt{N}\diag(\Xfd)\bar\Fm$. For channel estimation in OFDM, $Q$ equally spaced pilot signals are inserted at the transmitter \cite{cai2004error,islam2011optimum,al2010model}. Based on this known pilot sequence, the receiver finds the MMSE estimate of the channel. Let $\Ic_q$ denote the index set of the pilot locations, then we can write $\Yfd_{\Ic_q}=\Xm_{\Ic_q}\hv+\Zfd^\prime_{\Ic_q}$, where $\Ufd_{\Ic_q}$ prunes $\Ufd$ of all rows except for the rows belonging to $\Ic_q$. Now the MMSE estimate of $\hv$ can be obtained by solving the regularized LS problem, $\hat\hv=\underset\hv{\arg\max}\{\|\Yfd_{\Ic_q}-\Xm_{\Ic_q}\hv\|_{\Rm_{\Zc^\prime}^{-1}}^2+\|\hv\|_{\Rm_h^{-1}}^2\}$ where $\Rm_h=\EE[\hv\hv^\herm]=\sigma_h^2\Id$. Further, by ignoring the clipping noise component of $\Zfd^\prime$, we can write $\Rm_{\Zc^\prime}=\EE[\Zfd^\prime{\Zfd^\prime}^\herm]=\sigma_z^2\Id$ (the subscript $\Ic_q$ of $\Zfd^\prime$ is dropped here for notational convenience). Solving this LS problem yields \cite{SayedBook}
\begin{align}\label{eqMMSEsimple}
\hat\hv=\Xm_{\Ic_q}^\herm(\Xm_{\Ic_q}\Xm_{\Ic_q}^\herm+(\sigma_z^2/\sigma_h^2)\Id)^{-1}\Yfd_{\Ic_q}.
\end{align}

Increasing the number of pilot tones for CIR estimation results in improved estimation accuracy. However, generally it is not feasible to spare additional pilots as it reduces the data rate. In this work, we increase the number of measurements without increasing the number of reserved pilots by using RCs (for the procedure to find the RCs see the discussion following (\ref{eqrecsigfreqeq})). Let $\Ic_r$ denote the index set of the RCs and the pilot carriers. We can now retain these carriers in estimating $\hv$ and prune all other sub-carriers from (\ref{eqrecsigfreq3}). This yields
\begin{align}\label{eqrelcarr}
\Yfd_{\Ic_r}=\Xm_{\Ic_r}\hv+\Zfd_{\Ic_{r}}^\prime,
\end{align}
Now, we can obtain the refined estimate of $\hv$ based on (\ref{eqMMSEsimple}) by replacing the pilot index set $\Ic_q$ with enhanced set $\Ic_r$ consisting of the pilots and RCs. The enhanced MMSE estimation procedure based on RCs can be summarized in the following three steps: 1) Find the initial MMSE estimate of the CIR using (\ref{eqMMSEsimple}), 2) Find reliability $\mathfrak{R}$ for all sub-carriers using (\ref{eqreliability}) and select $R$ sub-carriers with the highest reliability index as RCs and 3) Use the RCs as additional measurements (by using (\ref{eqrelcarr})) and find MMSE estimate using (\ref{eqMMSEsimple}).

It is important to note that however many pilots and RCs we use to enhance the channel estimate, we are bottle-necked by the clipping distortion. Another way to look at this is to notice that what passes through the channel is not the pure signal or pilots but their clipped versions. As such, motivated by the work of \cite{wu2005iterative}, we first estimate the contaminated (pilots + RCs) and use them for enhanced MMSE estimation. The proposed data-aided CIR estimation scheme can be summarized as:
\begin{enumerate}
\item Obtain the initial MMSE estimate by using (\ref{eqMMSEsimple}).
\item Equalize the received data and make ML decisions on the equalized data i.e., $\lfloor\Yc(i)/\hat\Dc(i)\rfloor=\lfloor\hat\Xc(i)\rfloor$.
\item Find reliability $\mathfrak{R}$ for all sub-carriers and select $R$ sub-carriers with the highest reliability index as RCs.
\item Construct the time-domain signal $\hat\xv=\Fm^\herm\lfloor\hat\Xc(i)\rfloor$.
\item Find $\hat\xv_p$ by clipping $\hat\xv$ using (\ref{eqclip1}) and obtain $\hat\Xfd_p=\Fm\hat\xv_p$.
\item Obtain (clipped pilot sequence + RCs) $\Xfd_{p_{\Ic_r}}$ and $\Xm_{p_{\Ic_r}}=\diag(\Xfd_{p_{\Ic_r}})$.
\item Use $\Xm_{p_{\Ic_r}}$ in (\ref{eqMMSEsimple}) to obtain the improved CIR estimate.
\end{enumerate}
\subsection{Simulation Results}
\begin{figure}[h!]\centering
\includegraphics[width=0.45\textwidth]{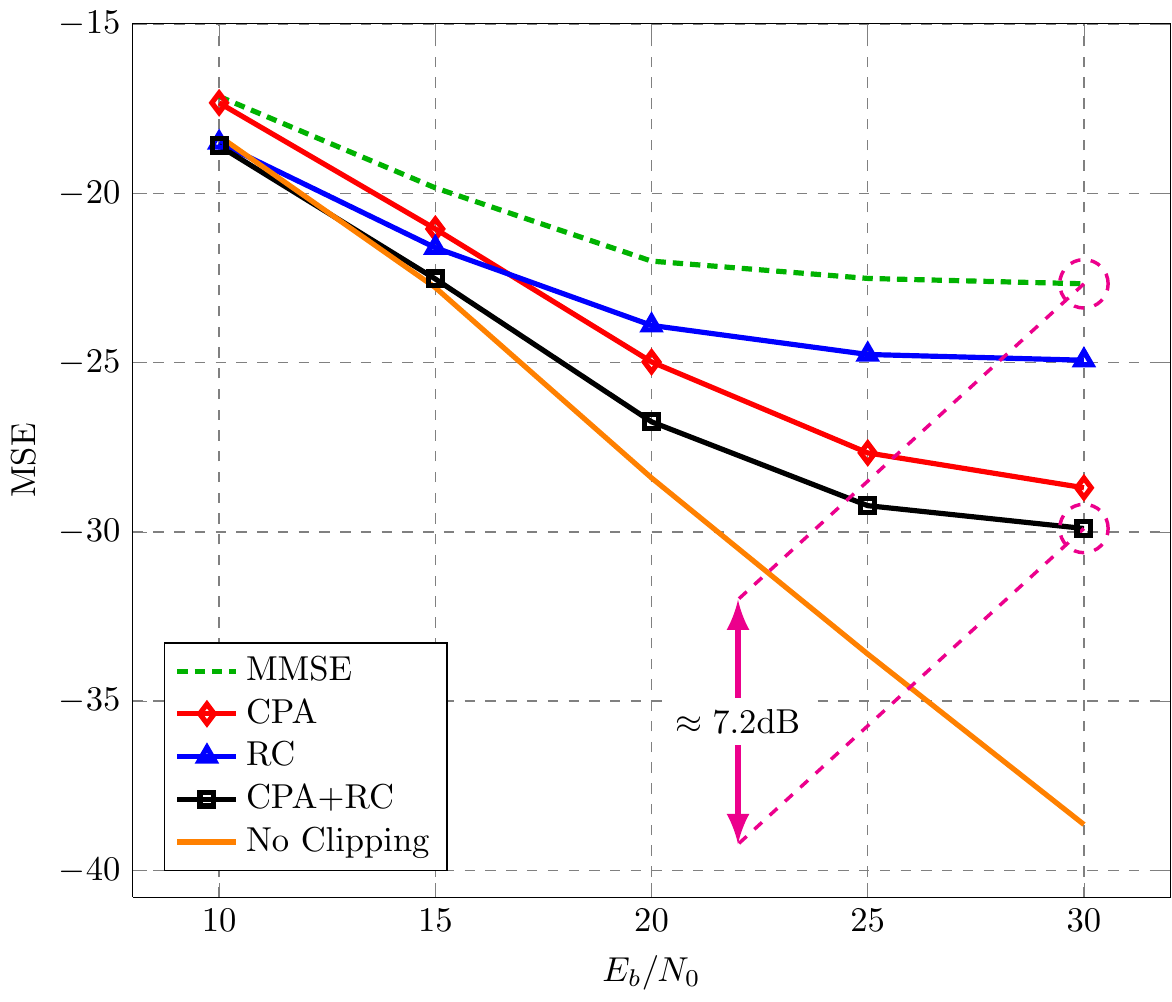}
\caption{$E_b/N_0$ MSE (dB) for data-aided CIR Estimation ($\mathrm{CR}=1.73,Q=R=16$).}
\label{fig:ChEst}
\end{figure}

\begin{figure}[h!]\centering
\includegraphics[width=0.45\textwidth]{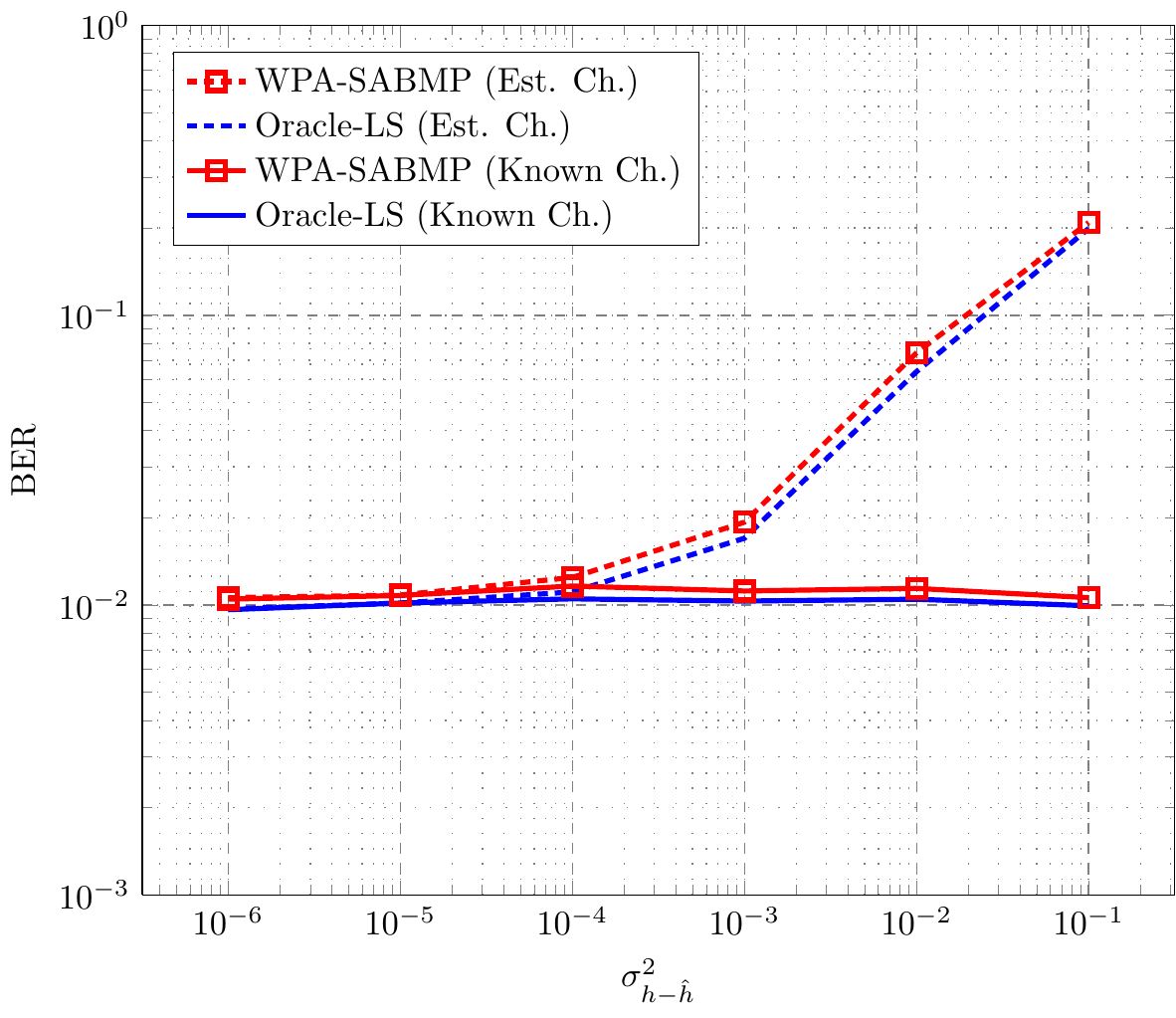}
\caption{BER performance of the proposed scheme as a function of the channel estimation error ($\mathrm{CR}=1.62,E_b/N_0=20$dB).}
\label{fig:ChEst2}
\end{figure}

For channel estimation $256$ sub-carrier OFDM and $64$-QAM modulation is used. A total of $16$ equispaced pilots are inserted to estimate a rayleigh fading channel of length $10$. For estimation the number of RCs is chosen to be $16$ (i.e., $Q=R=16$). Fig.~\ref{fig:ChEst} shows the mean squared error (MSE) results of simple MMSE estimation (MMSE), the RCs approach (RC), the contaminated pilot approach (CPA) \cite{wu2005iterative}, the proposed scheme (RC+CPA) and the MMSE for unclipped OFDM (No clipping). The MSE as a function of $E_b/N_0$ results are generated by keeping $\mathrm{CR}=1.73$. The results show that for high $E_b/N_0$ the proposed scheme provides upto $7.2$ dB advantage over simple MMSE estimation. Finally, we test the performance of the proposed scheme under channel estimation error. The results of this experiment are plotted as a function of channel error variance $\sigma_{h-\hat h}^2$ and are shown in Fig.~\ref{fig:ChEst2}. It can be observed from the results, that the performance of the proposed scheme is slightly affected for $\sigma_{h-\hat h}^2$ upto $10^{-3}$ and deteriorates only after $\sigma_{h-\hat h}^2$ exceeds $10^{-3}$.
\section{Conclusion}\label{secConc}
In this work, a low complexity Bayesian clipping recovery scheme was presented. The proposed WPA-SABMP scheme utilizes the undistorted phase property and weighting for enhanced clipping recovery. The proposed approach is agnostic to the non-Gaussian distribution of the clipping signal and thus outperforms other traditional Bayesian approaches and $\ell_1$-sparse recovery schemes. The WPA-SABMP scheme also utilizes the available statistics for enhanced recovery, however, when these statistics were unavailable the proposed scheme bootstrapped itself and successfully estimated the clipping distortion. Simulation results showed significant performance enhancement for WPA-SABMP scheme in both the error rate and complexity. The proposed scheme was then extended for the SIMO-OFDM systems and numerical findings were presented. In addition, a multi-user clipping recovery scheme was proposed and channel estimation strategies were presented for clipped OFDM signal. The simulation results for OFDMA clipping mitigation and data-aided channel estimation also showed favorable results.
\bibliographystyle{IEEEtran}
\bibliography{WPA_SABMP_Access_2014}
\end{document}